\documentclass[reprint,amsmath,amssymb,aps,pre]{revtex4-1}  

\usepackage{graphicx}
\usepackage{dcolumn}
\usepackage{bm}
\usepackage{hyperref}
\hypersetup{colorlinks=True,citecolor=blue,linkcolor=blue,urlcolor=blue}
\usepackage{multirow} 
\usepackage{enumerate}

\def\be{\begin{equation}} 
\def\ee{\end{equation}} 
\def\intd{\,\mathrm{d}} 

\begin{document}

\title{CHNS: A Case Study of Turbulence in Elastic Media}
\author{Xiang Fan}\affiliation{University of California at San Diego, La Jolla, California 92093}
\author{P. H. Diamond}\affiliation{University of California at San Diego, La Jolla, California 92093}
\author{L. Chac\'on}\affiliation{Los Alamos National Laboratory, Los Alamos, New Mexico 87545}
\date{\today} 

\begin{abstract}
Recent progress in the study of Cahn-Hilliard Navier-Stokes (CHNS) turbulence is summarized. This is an example of \textit{elastic turbulence}, which can occur in elastic (i.e. self-restoring) media. Such media exhibit memory due freezing-in laws, as does MHD, which in turn constrains the dynamics. We report new results in the theory of CHNS turbulence in 2D, with special emphasis on the role of structure (i.e. ``blob'') formation and its interaction with the dual cascade. The evolution of a concentration gradient in response to a single eddy -- analogous to flux expulsion in MHD -- is analyzed. Lessons learned are discussed in the context of MHD and other elastic media.
\end{abstract}

\maketitle
\section{Introduction}

The study of \textit{active scalar} turbulence is a central focus of research in theoretical plasma physics. Active scalar systems are ubiquitous. Examples include turbulent flow with polymer additives (for drag reduction), flow with bubbles, strong surface waves in the presence of surfactants, etc. Examples in plasma physics include 2D MHD and reduced MHD \cite{seshasayanan_edge_2014,seshasayanan_critical_2016,kim_are_2002,biskamp_magnetohydrodynamic_2003,biskamp_two-dimensional_2001,kraichnan_inertial-range_1965,iroshnikov_turbulence_1964,pouquet_strong_1976,pouquet_two-dimensional_1978,fyfe_high-beta_1976,gruzinov_self-consistent_1995,matthaeus_rapid_2008,servidio_magnetic_2009,diamond_self-consistent_2005,diamond_modern_2010,frisch_possibility_1975,celani_active_2002,banerjee_statistics_2014}, and their generalizations to include ballooning coupling \cite{connor_two_1985,hastie_drift_2003}, the Hasegawa-Wakatani model and related fluid systems for drift wave and ITG turbulence \cite{hasegawa_pseudothreedimensional_1978,hasegawa_plasma_1983,li_negative_2017,li_dynamics_2016,mattor_momentum_1988}, as well as many other examples. Active scalar systems are logical outcomes of the model reduction process used to simplify the full 2-fluid Braginsky system \cite{braginskii1965reviews} in the case of strong magnetization and weakly compressible dynamics (i.e. which excludes magnetosonic time scales). Of course, active scalar problems are to be contrasted to the familiar case of a \textit{passive scalar}, in that they involve feedback of the advected fields on the fluid dynamics. Strongly magnetized active scalar problems have the generic structure of:

\begin{enumerate}
\item A vorticity equation, with linear and/or nonlinear couplings to the advected scalar. This follows from $\nabla\cdot\mathbf{J}=0$ with $\nabla\cdot\mathbf{J}_{pol}=-\nabla_{\parallel}\cdot\mathbf{J}_{\parallel}-\nabla_{\perp}\cdot\mathbf{J}_{PS}$. Here $\mathbf{J}_{pol}$ is the perpendicular polarization current, $\mathbf{J}_{\parallel}$ is the current parallel to $\mathbf{B}_0$, and $\mathbf{J}_{PS}$ is the Pfirsch-Schluter current, related to curvature. For $\mathbf{E}\times\mathbf{B}$ velocity, $\nabla\cdot\mathbf{J}_{pol}$ reduces to vorticity evolution. Note that all reduced fluid models contain an equation of this form, as $\nabla\cdot\mathbf{J}=0$ is fundamental.
\item A scalar advection equation. For 2D MHD, this is simply the statement of conservation of magnetic potential $A$. For the Hasegawa-Wakatani system, it is the density equation, which involves linear coupling of density and potential.
\end{enumerate}

Many reduced active scalar systems exhibit elasticity - i.e. the tendency of the flow to be self-restoring (i.e. ``springy'') -- due to memory enforced by a freezing-in constraint. 2D MHD is a prime example of an elastic active scalar system, in which the ``springiness'' is due to magnetic tension, and memory follows from Alfv\'en's Theorem. 

\begin{figure}[htbp] 
    \centering
    \includegraphics[width=\columnwidth]{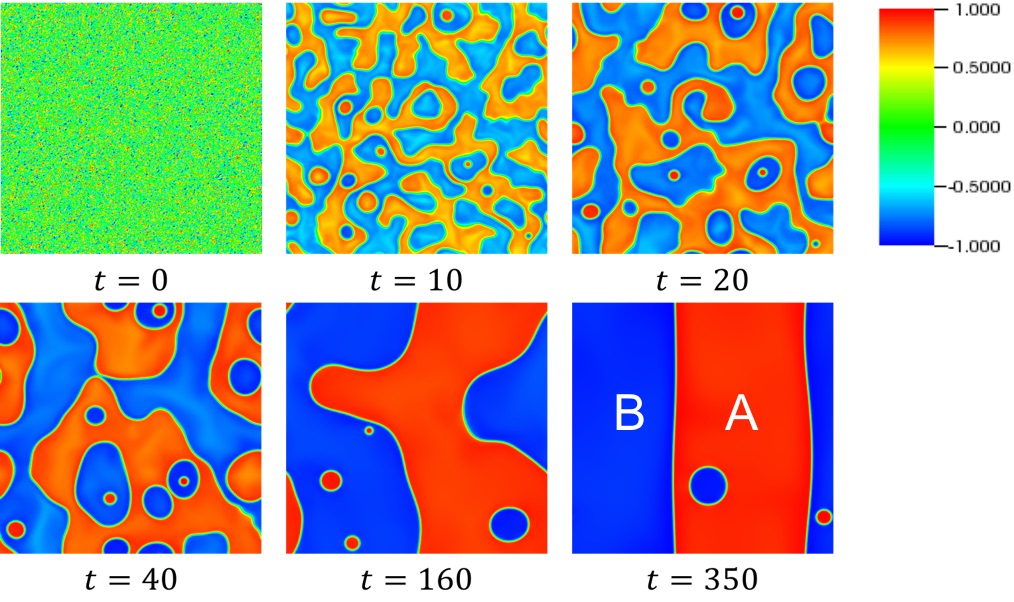}
    \caption{$\psi$ field evolution as an illustration of spinodal decomposition. The red and the blue are colors for the two components of the binary fluid.}
    \label{spinodal_decomposition}
\end{figure}

The Cahn-Hilliard Navier-Stokes (CHNS) model in 2D is an active scalar system with many interesting similarities to, and differences from, 2D MHD and other active scalar systems relevant to MFE physics \cite{fan_cascades_2016,fan_formation_2017,ruiz_turbulence_1981,pandit_overview_2017}. The CHNS system describes the motion and evolution of phase separation (spinodal decomposition) of two immiscible fluids \cite{cahn_free_1958,cahn_spinodal_1961,perlekar_two-dimensional_2017,perlekar_droplet_2012,perlekar_spinodal_2014,pal_binary-fluid_2016,datt_morphological_2015,tierra_numerical_2014,guillen-gonzalez_second_2014,furukawa_spinodal_2000,ernst_observation_1992,siggia_late_1979,kendon_inertial_2001,kendon_3d_1999,berti_turbulence_2005,berthier_phase_2001,lacasta_phase_1995,o_naraigh_bubbles_2007,pine_turbulent_1984,huang_study_1974,guan_second_2014,liu_isogeometric_2013,kim_conservative_2004}. The CHNS system has applications to alloy, cell sorting and other dynamic phase separation phenomena. See Fig.~\ref{spinodal_decomposition} as an illustration. The 2D CHNS equations are:

\be
\partial_t\psi+\mathbf{v}\cdot\nabla\psi=D\nabla^2(-\psi+\psi^3-\xi^2\nabla^2\psi)\label{CHNS1}
\ee
where the scalar field $\psi(\mathbf{r},t)\equiv (\rho_A(\mathbf{r},t)-\rho_B(\mathbf{r},t))/\rho$ is the normalized component density contrast, and:

\be
\partial_t\omega+\mathbf{v}\cdot\nabla\omega=\frac{\xi^2}{\rho}\mathbf{B}_\psi\cdot\nabla\nabla^2\psi+\nu\nabla^2\omega \label{CHNS2}\\
\ee
where $\omega$ is the vorticity. Here $\mathbf{v}=\mathbf{\hat{z}}\times\nabla\phi\label{CHNS3}$ defines a scalar potential and $\mathbf{B}_\psi=\mathbf{\hat{z}}\times\nabla\psi$. Also, $\nu$ is the viscosity, $D$ is the scalar diffusivity, and $\xi$ is a parameter characterizing the width of the interface between ``blobs'' of phases A and B. Note that $\psi$ takes on values only in the range $-1\leq \psi\leq 1$ by definition. The similarities and differences of the 2D CHNS system and 2D MHD (the prototype of magnetized active scalar systems) are evident. Note also that the negative diffusivity in the scalar advection equation assures the formation of clusters or ``blobs'' of $\psi\rightarrow +1$ and $\psi\rightarrow -1$ phase domains in the system. The aim of this paper is to elucidate the physics of active scalar turbulence by the study of the new (to the plasma community) CHNS system, which manifests both classic themes and new twists in active scalar turbulence. A second aim of this paper is to extract the more general lessons learned from this work and to indicate where they might be applied to more familiar models of plasma turbulence.

Active scalar models present several challenges, the resolution of which are important for understanding the multi-field turbulence in such systems. Three prominent physics issues here are: (1) dual or multiple cascades, (2) the nature of ``blobby'' turbulence and the scale selection problem inherent to it, and (3) negative diffusion and up-gradient transport. They are discussed below.

Regarding cascades, our present understanding of plasma turbulence is strongest for single equation / single field models, like the Hasegawa-Wakatani system, or ITG with Boltzmann electrons. Even MHD presents new problems, such as which cascade is ``fundamental''?-- i.e. the inverse $\langle A^2\rangle$ (2D) or magnetic helicity $\langle \mathbf{A}\cdot\mathbf{B}\rangle$ (3D) cascade, or the forward energy cascade. The theoretical focus is primarily on the inertial range for the latter, due to its being the natural extension of the archetypical Kolmogorov cascade in the 3D Navier-Stokes problem. However, the inverse magnetic cascades are closely related to the freezing-in law, which exerts fundamental topological constraints, and so are \textit{at least} of equal importance. Note that virtually \textit{all} models of electro-magnetic turbulence in magnetized plasmas are built upon the foundation of 2D MHD and its close relative, reduced MHD. Thus all such systems will support dual (or multiple!) cascades, and so present to us questions like those posed above. 

``Blobby'' turbulence refers to turbulence in which a gas or ``soup'' of structures forms and influences the dynamics and transport. Blobby turbulence is of great importance to SOL and divertor physics \cite{boedo_transport_2003,fuchert_characterization_2016,dippolito_convective_2011}. Indeed, the SOL density fluctuation PDF manifests a striking positive skewness, suggesting that $\tilde{n}>0$ structures are somehow preferred, and are a significant component of the turbulence. However, despite an uncountable number of impressive color view graphs devoted to this subject, there is little understanding of ``what makes a blob a blob'', i.e. what sets the scale of a blob, or how the blobs co-exist with, and influence, cascades.

Negative diffusion (i.e. ``negative viscosity'') phenomena and up-gradient transport are processes fundamental to the formation of macroscopic flows in turbulence. Zonal flow formation is a particularly important negative viscosity phenomenon in magnetized plasmas \cite{diamond_zonal_2005,hasegawa_self-organization_1987}. Indeed, zonal flow formation closely resembles the process of phase separation or spinodal decomposition, in which a mixture separates into domains of different components. In this context, zonal flow formation may be thought of as a spinodal decomposition of a mixture of fluid elements with poloidal $\mathbf{E}\times\mathbf{B}$ flow velocity $>0$ or $<0$ (but equal in magnitude) into neighboring bands (domains), with net momentum $>0$ and $<0$, respectively. Such a decomposition requires up-gradient momentum transport, and so is a type of ``negative viscosity'' process.

This paper argues that CHNS turbulence illuminates all three of the challenges listed above. CHNS turbulence exhibits a dual cascade, where energy is scattered forward, while $\langle\psi^2\rangle$ undergoes inverse transfer. We shall see, though, that the \textit{inverse} cascade process is more robust, and actually tends to alter the forward cascade. CHNS is also elastic, due to surface tension restoring forces, and this elasticity is the physical process underpinning of the analogy between CHNS and MHD. In addition, CHNS is intrinsically ``blobby'', and structures form and grow in time according to $l\sim t^\alpha$, $\alpha\sim 2/3$. However, we show that blob coalescence actually reduces the region of elastic feedback on the flow, and so modifies the cascade and the energy spectrum power law. Finally, negative viscosity phenomena are central to the CHNS system, which is a prototype for spinodal decomposition. Scale selection in CHNS (for the Hinze scale) occurs by the competition between blob growth and fluid straining. Thus, the CHNS system offers a good opportunity to understand the detailed physics of negative diffusion processes and how to regulate them.

In this paper we discuss and review computational and theoretical progress toward understanding 2D CHNS turbulence and single eddy mixing, with the aim of extracting general insights concerning elastic active scalar systems in magnetized plasmas. As suggested by equilibrium statistical mechanics, 2D CHNS, with vorticity forced at large scales, manifests a dual cascade of energy forward, $\langle\psi^2\rangle$ inverse. The eddy transfer couples to elastic waves, and an elastic range forms from the emergent Hinze scale (where Reynolds and elastic stresses balance) down to small scale dissipation. The inverse cascade of $\langle\psi^2\rangle$ is closely related to the real space dynamics of blob formation and merger. Interestingly, while the spectral exponent for $\langle\psi^2\rangle_k$ follows standard scaling predictions, that for the energy spectrum does not. We resolve this apparent puzzle by observing that as blob merger progresses, thus forming larger blobs, the extent of surface tension feedback on the flow decreases. This is because the effective extent of $|\nabla\psi|\neq 0$ regions declines as blobs coalesce, thus reducing the ``active region'' for feedback. Hence, the evolution of the vorticity more closely resembles that of a simple 2D fluid.

We also discuss single eddy mixing, motivated by the tendency of blobs to coalesce to form large structures studies. These studies explore the evolution of a $|\nabla\psi|\neq 0$ layer in the presence of a single, sheared eddy, with negative diffusion, positive hyper-diffusion and dissipative nonlinearity. The study of this system, which is the CHNS analogue of the classic MHD paradigm of flux expulsion, offers basic insights into mixing and the interaction of flow shear with the dissipative evolution of the scalar field $\psi$. The characteristic hybrid time scales for mixing are determined, and the multi-stage evolution of mixing is elucidated. An interesting outcome of this study is the observation that long lived target patterns form in the $\psi$ field. These exhibit progressive pairwise mergers on an exponentially long time scale, as do ``steps'' in staircase layering structures. 

The remainder of this paper is organized as follows. Section~\ref{MHD_section} presents a comparison and contrast of 2D CHNS and 2D MHD, two active scalar systems. Section~\ref{CHNS_section} discusses CHNS turbulence as a case study in elastic active scalar turbulence. We emphasize the complementary k-space and real space description of the evolution. The mixing of layers by a single eddy is analyzed in Section~\ref{target_section}. The formation of long lived target patterns is observed and discussed. Section~\ref{conclusion_section} presents conclusions, plans for future work, and discusses the broader lessons learned in this research.

\section{2D MHD and 2D CHNS: Comparing and Contrasting}\label{MHD_section}

The 2D MHD system is:
\begin{align}
\partial_t A+\mathbf{v}\cdot\nabla A&=\eta\nabla^2 A \label{MHD1}\\
\partial_t\omega+\mathbf{v}\cdot\nabla\omega&=\frac{1}{\mu_0\rho}\mathbf{B}\cdot\nabla\nabla^2 A+\nu\nabla^2\omega \label{MHD2}
\end{align}
where $A$ is the scalar magnetic potential, $\mathbf{B}=\mathbf{\hat{z}}\times\nabla A$ is the magnetic field, $\eta$ is the resistivity, and $\mu_0$ is magnetic permeability. The 2D CHNS equations Eq.~(\ref{CHNS1}-\ref{CHNS2}) are mentioned in the introduction. The origin of these equations is the Landau theory for second order phase transition. The order parameter is the concentration $\psi$, and the free energy is
\be
F[\psi]=\int (-\frac{1}{2}\psi^2+\frac{1}{4}\psi^4+\frac{\xi^2}{2}|\nabla\psi|^2)\intd\mathbf{r}
\ee
A graph of the first two terms has a ``W'' shape, with minimums at $\psi=\pm1$. Thus the system tends to undergo a phase separation process given a small initial perturbation around $\psi=0$. The CHNS equations do not prevent the values of $\psi$ to go beyond $[-1, +1]$; however, the ``W''-shaped free energy confines the value of $\psi$ within $[-1, +1]$, without special numerical treatment. Of course, the structure of the ``W’’ curve implies that $\psi$ is attracted to the two minima at $\psi=\pm 1$. The third term is a curvature penalty, and it means that the $\psi$ field tends to have zero gradient inside blobs and to minimize the length of blob interfacial layers. The chemical potential is then given by $\mu=\delta F/\delta\psi=-\psi+\psi^3-\xi^2\nabla^2\psi$. Combine it with Fick's Law $\mathbf{J}=-D\nabla\mu$ and continuity equation $\mathrm{d}\psi/\mathrm{d}t+\nabla\cdot\mathbf{J}=0$, it is straightforward to obtain the CHNS equations. As shown in Fig.~\ref{spinodal_decomposition}, the blobs in the CHNS system tend to aggregate. If the system is unforced, the coalescence process will continue until the blob size reaches the system size. 

The comparison and contrast of the basic elements of the two systems (2D MHD and CHNS) are summarized in Table.~\ref{comparison}, and the contents will be explained in this and the next sections. Comparing the 2D MHD system Eq.~(\ref{MHD1}-\ref{MHD2}) and the 2D CHNS system Eq.~(\ref{CHNS1}-\ref{CHNS2}), it is easy to find that both sets of equations contain an evolution equation for a scalar field and a vorticity equation. The magnetic potential $A$ in MHD is analogous to the concentration field $\psi$ in CHNS. Other analogues are shown in the Table~\ref{correspondence}. The back reaction terms from the scalar field on the fluid motion have the same form, up to a change of variable. $\frac{1}{\mu_0\rho}\mathbf{B}\cdot\nabla\nabla^2 A$ is due to the $\mathbf{j}\times\mathbf{B}$ force in 2D MHD,  $\frac{\xi^2}{\rho}\mathbf{B}_\psi\cdot\nabla\nabla^2\psi$ is the surface tension force in 2D CHNS. The difference between these two systems is in the dissipative part of the scalar evolution equation. In 2D MHD, there is only a simple diffusion of $A$; however, in 2D CHNS, the dissipative terms are more complicated. CHNS has a negative diffusion term, a dissipative self nonlinear term, and a hyper-diffusion term.

\begin{table}
\caption{The correspondence between 2D MHD and the 2D CHNS system.}
\begin{center}
\begin{tabular}{ccc}
\hline
\hline
& 2D MHD & 2D CHNS \\
\hline
Magnetic Potential & $A$ & $\psi$ \\
Magnetic Field & $\mathbf{B}$ & $\mathbf{B}_\psi$ \\
Current & $j$ & $j_\psi$ \\
Diffusivity & $\eta$ & $D$ \\
Interaction strength & $\frac{1}{\mu_0}$ & $\xi^2$ \\
\hline
\hline
\end{tabular}
\end{center}
\label{correspondence}
\end{table}%

The CHNS system supports a linear elastic wave, and its dispersion relation is:
\be
\omega(\mathbf{k})=\pm\sqrt{\frac{\xi^2}{\rho}}|\nabla\psi_0\times\mathbf{k}|-\frac12i(CD+\nu)k^2
\ee
where $C=[-1-6\psi_0\nabla^2\psi_0/k^2-6(\nabla\psi_0)^2/k^2-12\psi_0\nabla\psi_0\cdot i\mathbf{k}/k^2+3\psi_0^2+\xi^2k^2]$ is a dimensionless coefficient. This wave is similar to a capillary wave at an interface between two fluids, because surface tension generates the restoring force. It only propagates along the interfaces between the blobs, where $|\mathbf{B}_{\psi}|$ is nonzero. This wave is also analogous to an Alfv\'en wave in 2D MHD. The two waves have similar dispersion relation, and they both propagate along $\mathbf{B}_{\psi}$ or $\mathbf{B}$ field lines. Both surface tension and magnetic field act as elastic restoring forces. Besides, the linear elastic wave leads to elastic equipartition, and it further affects the spectra power law in the same way as the Alfv\'enic equipartition does in 2D MHD, as discussed later in this paper. There are also important differences. The $\mathbf{B}_{\psi}$ field in CHNS is large only in the inter-facial regions, but the magnetic field in MHD can be significant everywhere. Therefore, the elastic wave activity in CHNS does not fill the whole space, while Alfv\'enic feedback does.

\begin{table*}
\caption{Comparison and contrast of 2D MHD and the 2D CHNS system.}
\scriptsize
\begin{center}
\begin{tabular}{ccc}
\hline
\hline
& 2D MHD & 2D CHNS\\
\hline
Diffusion & A simple positive diffusion term & A negative, a self nonlinear, and a hyper-diffusion term\\
Range of potential & No restriction for range of $A$ & $\psi\in[-1,1]$ \\
Origin of elasticity & Magnetic field induces elasticity & Surface tension induces elasticity\\
Waves & Alfv\'en wave & CHNS linear elastic wave\\
Ideal Quadratic Conserved Quantities & Conservation of $E$, $H^A$ and $H^C$ & Conservation of $E$, $H^\psi$ and $H^C$\\
The inverse cascades & Inverse cascade of $H^A$ & Inverse cascade of $H^\psi$\\
Origin of the inverse cascades & The coalescence of magnetic flux blobs & The coalescence of blobs of the same species\\
Inverse cascade spectra & $H^A_k\sim k^{-7/3}$ & $H^\psi_k\sim k^{-7/3}$\\
The forward cascades & Suggestive of direct energy cascade & Suggestive of direct enstrophy cascade\\
Kinetic energy spectra & $E^K_k\sim k^{-3/2}$ & $E^K_k\sim k^{-3}$\\
Interface Packing Fraction & Not far from $50\%$ & Small\\
Back reaction & $\mathbf{j}\times\mathbf{B}$ force can be significant & Back reaction is apparently limited\\
\hline
\hline
\end{tabular}
\end{center}
\label{comparison}
\end{table*}

The ideal quadratic conserved quantities in 2D CHNS are the same as these in 2D MHD, up to a change of variable. The difference between these two systems is only in the non-ideal terms. The three ideal quadratic conserved quantities are:
\begin{align}
E=E^K+E^B&\equiv\int(\frac{\rho\mathbf{v}^2}{2}+\frac{\xi^2\mathbf{B}_\psi^2}{2})\intd^2x \label{CHNS_energy}\\
H^\psi=\langle\psi^2\rangle&\equiv\int \psi^2\intd^2x\\
H^C&\equiv\int\mathbf{v}\cdot\mathbf{B}_\psi\intd^2x
\end{align}
where $E$ is energy, $E^K$ and $E^B$ are kinetic and elastic energy, respectively, $H^\psi$ is mean square concentration, and $H^C$ is cross helicity. $H^\psi$ is analogous to the mean square magnetic potential $H^A\equiv\int A^2\intd^2x$ in 2D MHD.

\section{CHNS: A case study in Active Scalar Turbulence}\label{CHNS_section}

\begin{figure}[htbp] 
    \centering
    \includegraphics[width=\columnwidth]{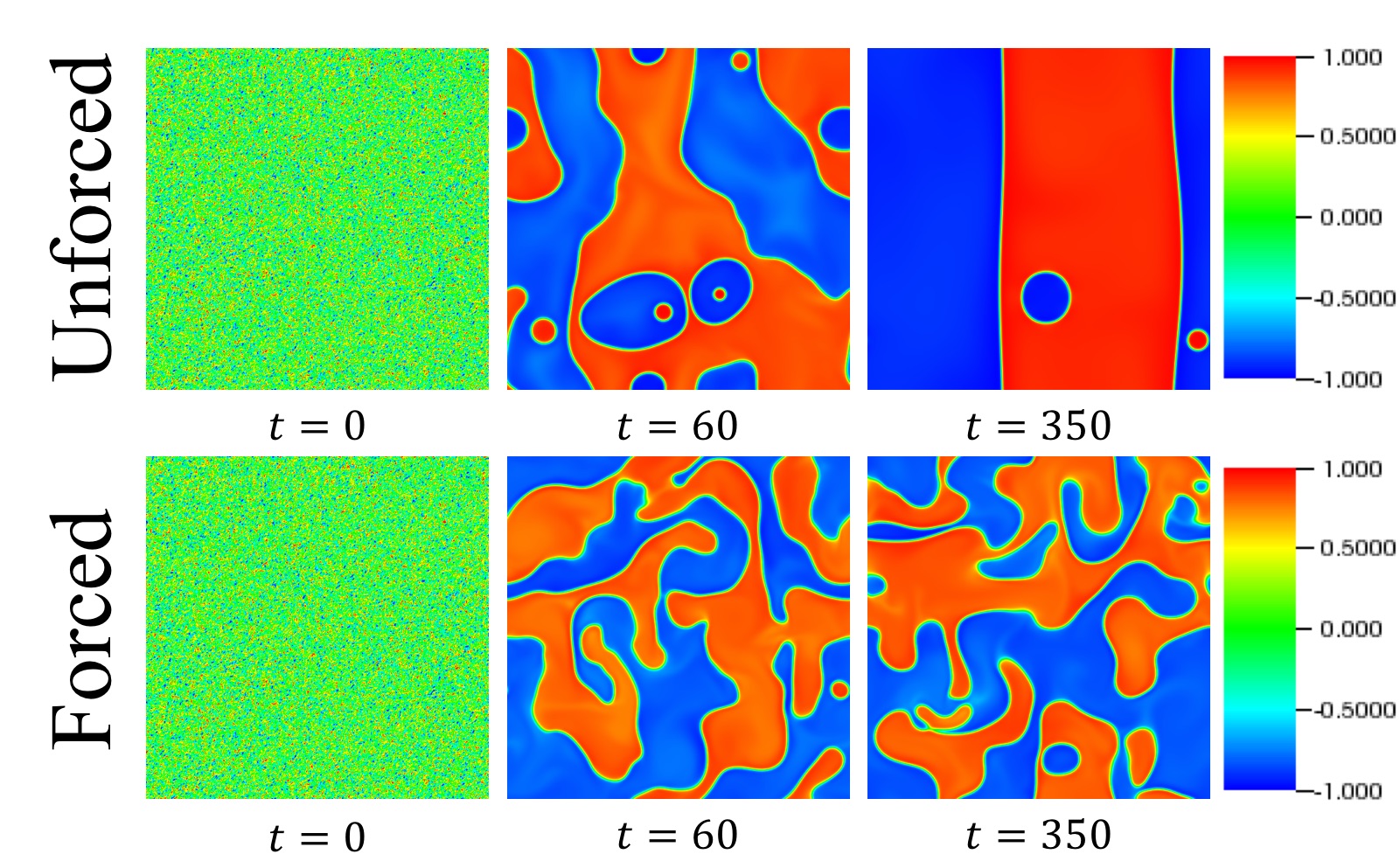}
    \caption{Top panels are $\psi$ field evolution plots for an unforced run at various times; bottom panels are the ones for a forced run. Reprinted with permission from Xiang Fan, P. H. Diamond, L. Chac\'on, and Hui Li, Phys. Rev. Fluids \textbf{1}, 054403 (2016). Copyright 2016 American Physical Society.}
    \label{forced}
\end{figure}

As shown in Fig.~\ref{spinodal_decomposition}, if the CHNS system is unforced, the blob size will grow continuously until it reaches the system size. More quantitatively, the length scale of the blobs grows as a power law of time $L\sim t^{2/3}$ \cite{furukawa_spinodal_2000,fan_cascades_2016}. If an external forcing at large scale is imposed on the vorticity field, then the large eddies will be broken up into smaller eddies. The blob coalescence process and fluid straining induced fragmentation will compete with each other. When they balance, blob size growth is arrested, and a statistically stable length scale for the blob size emerges. See Fig.~\ref{forced} as an illustration. The estimation of this final length scale is similar to the estimation of the typical size of a raindrop in a turbulent atmosphere. The raindrop size can be estimated by the balance of turbulent kinetic energy and surface tension energy. The scale at which these two balance defines the scale of a droplet, and it is called the Hinze scale \cite{hinze_fundamentals_1955,perlekar_spinodal_2014,perlekar_droplet_2012}. We can calculate the Hinze scale for the CHNS system as well by balancing the turbulent kinetic energy and the elastic energy. The result is:
\be
L_H\sim(\frac{\rho}{\xi})^{-1/3}\epsilon_\Omega^{-2/9}\label{Hinze_dir}
\ee

The range between the Hinze scale $L_H$ and the dissipation scale $L_d$ is defined to be the \textit{elastic range}. This is the range where kinetic and elastic energy are exchanged, and so elastic effects are significant. In this range, the dynamics is more MHD-like. $L_H\gg L_d$ is required for a long elastic range, and this is the case of interest. 

The blob coalescence process in the elastic range of the CHNS system is analogous to magnetic flux cell coalescence in MHD. Magnetic cell coalescence is the physical process which underlies the inverse cascade of mean square magnetic potential $H^A$ in 2D MHD. This suggests the inverse cascade of mean square concentration $H^\psi$ in 2D CHNS is due to hierarchical blob merger. This conclusion is also supported by statistical mechanics studies. Based on the ideal quadratic conserved quantities, absolute equilibrium distributions can be obtained. The real turbulent systems with finite dissipation are of course different from ideal systems, but the ideal conserved quantities still reflect important constraints on the nonlinear dynamics. The directions of the turbulent cascades are suggested. An inverse cascade of $H^\psi$ and a forward cascade of energy are expected for 2D CHNS, by analogy with 2D MHD. The inverse cascade of $H^\psi$ is a formal expression of the blob coalescence process. The forward cascade of energy is as usual, since the elastic force breaks enstrophy conservation.

Our simulation \cite{fan_cascades_2016,chacon_2d_2003,chacon_implicit_2002} also verified the inverse cascade of $H^\psi$, as shown in Fig.~9 (right) in Ref.~\cite{fan_cascades_2016}. In this simulation, there is no external forcing on $\psi$ field, and there is a homogeneous isotropic forcing at wave number $k=4$ on the $\phi$ field. The $H^\psi_k$ flux is defined to be $\Pi_{H\psi}(k)=\sum_{k<k'}T_{H\psi}(k')$, where $T_{H\psi}(k)=\langle \psi_k^*(\mathbf{v}\cdot\nabla \psi)_k\rangle$. The flux is negative, thus verified the direction of the cascade is indeed to large scales. In 2D MHD, the inverse cascade of $H^A$ is observed only if the $A$ field is perturbed at small scale. However, in 2D CHNS, a small scale perturbation of $\psi$ field is not necessary (for the inverse cascade of $H^\psi$) because fluctuations in $\psi$ to tend to aggregate.

The spectrum of $H^\psi$ exhibits a $k^{-7/3}$ power law, as shown in Fig.~10 (right) in Ref.~\cite{fan_cascades_2016}. This power law is the same as this for $H^A$ in 2D MHD. The derivation of this $k^{-7/3}$ power law is essentially the same for 2D MHD. The major assumptions are that: there is an elastic equipartition $\rho\langle v^2\rangle\sim\xi^2\langle B_\psi^2\rangle$ analogous to the Alfv\'enic equipartition; and that, the mean square magnetic potential spectral transfer rate $\epsilon_{H\psi}$ is constant. 

The $k^{-7/3}$ spectrum is robust. Different magnitudes of external forcing result in different Hinze scales, and thus in different extents of the elastic range. But within the elastic range, the power is still $k^{-7/3}$, as shown in Fig.~12 in Ref.~\cite{fan_cascades_2016}.

One may guess the kinetic energy power law for CHNS is $k^{-3/2}$, as is in 2D MHD. However, the actual power law is more close to $k^{-3}$, as shown in Fig.~13 (left) in Ref.~\cite{fan_cascades_2016}. Note that the energy power law for the 2D Navier-Stokes turbulence in the range of the forward enstrophy cascade is also $k^{-3}$ \cite{boffetta_two-dimensional_2012}. This suggests the back reaction of the $\psi$ field to the fluid motion is not as significant as for 2D MHD. 

\begin{figure}[htbp] 
    \centering
    \includegraphics[width=\columnwidth]{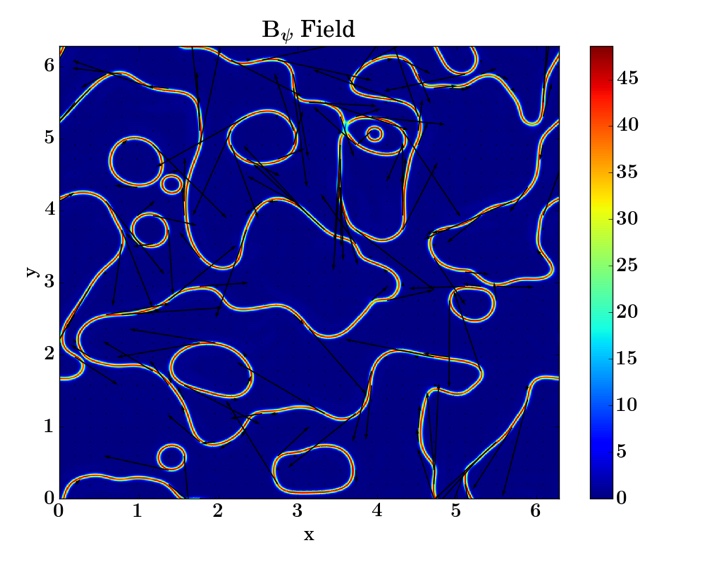}
    \caption{The $\mathbf{B}_\psi$ field for CHNS. Reprinted with permission from Xiang Fan, P. H. Diamond, L. Chac\'on, and Hui Li, Phys. Rev. Fluids \textbf{1}, 054403 (2016). Copyright 2016 American Physical Society.}
    \label{bpsi_field}
\end{figure}

\begin{figure*}[htbp] 
    \centering
    \includegraphics[width=\textwidth]{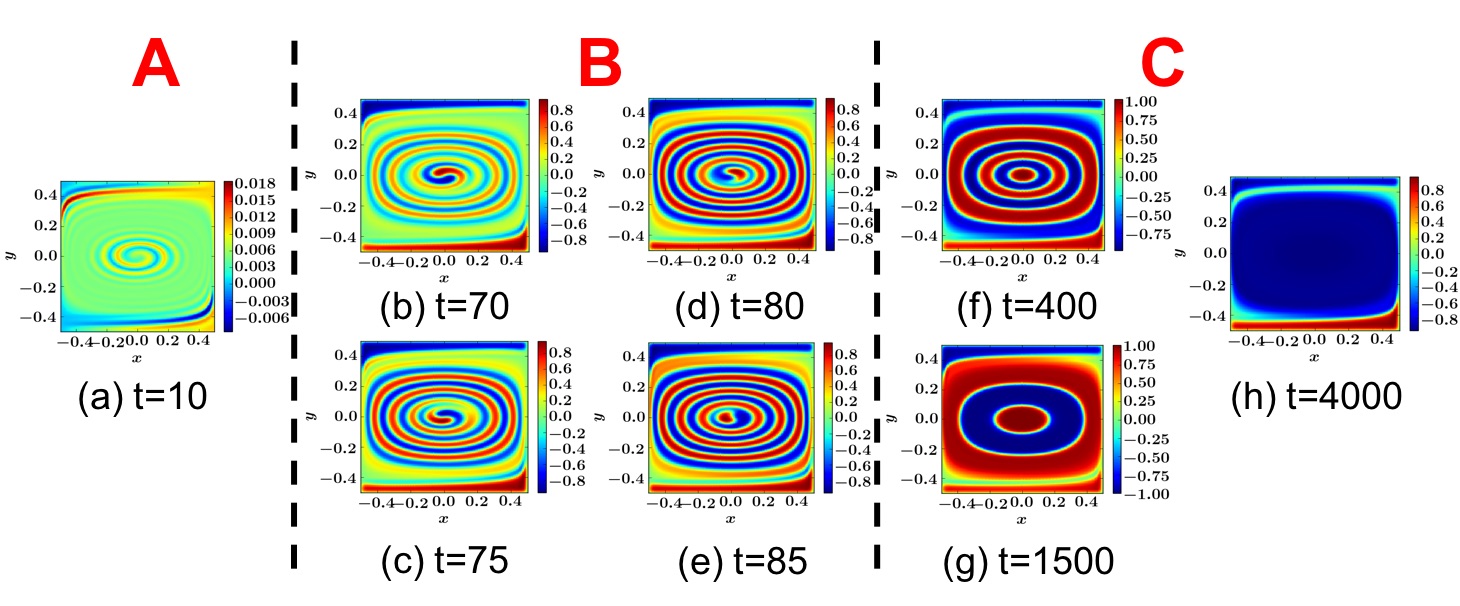}
    \caption{The evolution of the $\psi$ field. (a) The `jelly roll' stage, in which the stripes are spirals. (b) - (e) The topological evolution stage, in which the topology evolves from spirals to concentric annuli in the center of the pattern. (f) \& (g) The target pattern stage, in which the concentration field is composed of concentric annuli. The band merger progress occurs on exponentially long time scales; see (f) $\rightarrow$ (g) as an example. (h) The final steady state. Reprinted with permission from Xiang Fan, P. H. Diamond, and L. Chac\'on, Phys. Rev. E \textbf{96}, 041101(R) (2017). Copyright 2017 American Physical Society.}
    \label{psi_fields}
\end{figure*}

An obvious question now arises, which is very much the crux of the issue concerning 2D CHNS dynamics. This is: why does the CHNS $\leftrightarrow$ MHD correspondence apply so well for $H^\psi_k\sim H^A_k\sim k^{-7/3}$, yet break down drastically for energy? This initially surprising result can be understood by examining the \textit{real space structure} of the $\mathbf{B}_\psi$ field, as shown in Fig.~\ref{bpsi_field}. The distribution of $|\mathbf{B}_\psi|$ is significantly different in the regions of density contrast and inside the blobs. Note that elastic back-reaction in CHNS is limited to regions of density contrast, where $|\mathbf{B}_\psi|$ is significant. As blobs coalesce, the extent (i.e. length) of the A-B interfacial region decreases, so the ``Active region'' for elasticity drops as well. On the other hand, in MHD, magnetic fields pervade the system. More quantitatively, we define the interface packing fraction $P$ as:
\be
P\equiv\frac{\text{\# of grid points where }|\mathbf{B}_\psi|>B_\psi^{rms}}{\text{\# of grid points}}
\ee
Loosely put, $P$ may be thought of as the volume fraction where $|\mathbf{B}_\psi|$ is strong enough to generate appreciable elastic back reaction. As shown in Fig.~15 in Ref.~\cite{fan_cascades_2016}, $P$ for CHNS decays, while $P$ for MHD remains stationary. Smaller $P$ means a smaller region where the back reaction is significant, so the fluid dynamics is closer to simple Navier-Stokes. Therefore, the energy spectrum for CHNS resembles 2D Navier-Stokes turbulence more closely than it does 2D MHD turbulence.

All told, our study of 2D CHNS turbulence strongly suggests the importance of the study of real space structures on an equal footing with the traditional focus on k-space power laws.

\section{Relaxation and Mixing: A study of a Single Eddy}\label{target_section}

\begin{figure}[htbp] 
    \centering
    \includegraphics[width=\columnwidth]{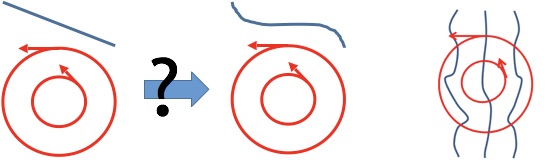}
    \caption{Understanding the competition of flow shearing and dissipation in the context of a \textit{single cell structure}.}
    \label{eddy}
\end{figure}

\begin{figure}[htbp] 
    \centering
    \includegraphics[width=\columnwidth]{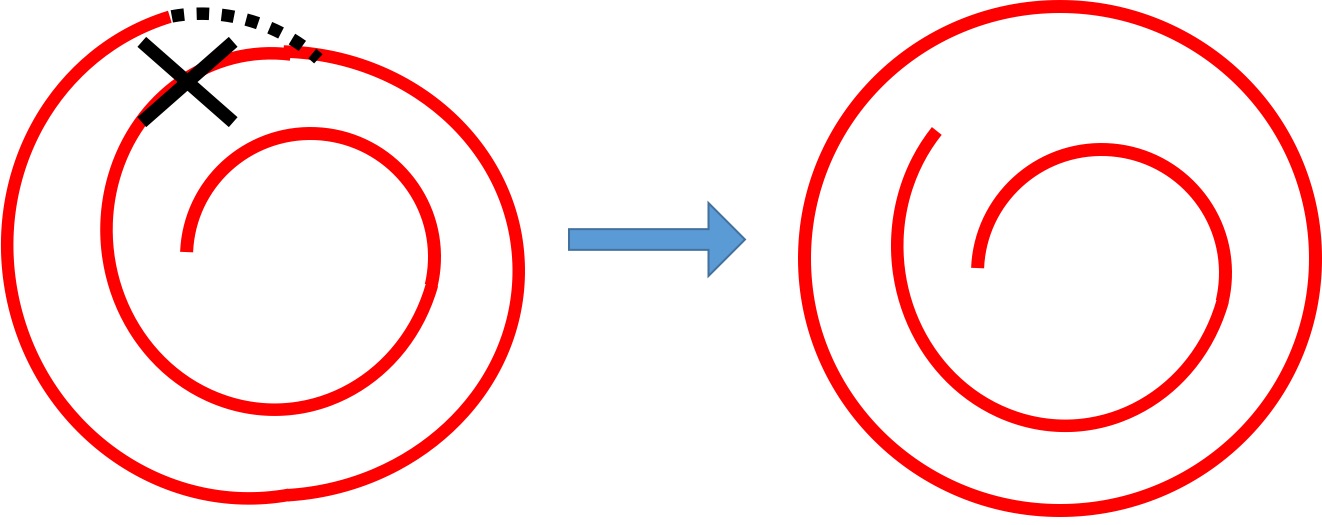}
    \caption{An illustration of the topological evolution from the jelly roll pattern to the target pattern: the stripes break in the middle, and the outer parts reconnect into a circle. Reprinted with permission from Xiang Fan, P. H. Diamond, and L. Chac\'on, Phys. Rev. E \textbf{96}, 041101(R) (2017). Copyright 2017 American Physical Society.}
    \label{topology_change}
\end{figure}

As discussed above, understanding real space structures evolution is the key to understand CHNS. Since the system tends to evolve to an end state of a few large blobs, the \textit{absolutely simplest} problem which emerges is that of understanding the competition of flow shearing and dissipation in the context of a \textit{single cell structure} (see Fig.~\ref{eddy}). This study of single eddy mixing in CHNS resembles the study of flux expulsion study in MHD \cite{gilbert_flux_2016,moffatt_time-scale_1983,moffatt_magnetic_1983,weiss_expulsion_1966}. The goals of the two studies are similar: to determine how the scalar field, and especially its spatial structure, evolves in the background of a fixed convective eddy. Also, the magnetic Reynolds number $\mathrm{Rm}$ and its analogue P\'eclet number $\mathrm{Pe}$ are $\gg1$ in both cases, so advection dominates. The analogous process for CHNS is the (kinematic) mixing of a region of $\nabla\langle\psi\rangle$ by a single, prescribed differentially rotating eddy in (dissipative) CHNS. The analogy with flux expulsion follows from the observation that $\mathbf{B}_{\psi0}=\nabla\langle\psi\rangle\times\mathbf{\hat{z}}$.

When a convective eddy is imposed in a weak magnetic field, the magnetic field is expelled and amplified outside the eddy. This MHD phenomena in MHD is called flux expulsion. Both simulation and analysis indicated that, the final value of $\langle B^2\rangle$ can be estimated by $\langle B^2\rangle\sim\mathrm{Rm}^{1/2}B_0^2$ where $\mathrm{Rm}$ is the magnetic Reynolds number, and the time for $\langle B^2\rangle$ to reach a steady state is $\tau_{MHD}\sim\mathrm{Rm}^{1/3}\tau_0$. Rhines and Young \cite{rhines_how_1983,rhines_homogenization_1982} noted the homogenization process (n.b. flux expulsion is closely related to PV homogenization in a 2D fluid) evolves through two stages: an initial rapid stage and a later slow stage. The rapid stage dynamics is dominated by shear-augmented diffusion, with time scale $\tau_{mix}\sim\mathrm{Rm}^{1/3}\tau_0$. The slow stage dynamics is simple diffusion, with time scale $\tau_{slow}\sim\mathrm{Rm}^1\tau_0$. However, in CHNS, the single eddy mixing exhibits more non-trivial evolutions.

In our simulation for the Cahn-Hilliard system \cite{fan_formation_2017,chacon_2d_2003,chacon_implicit_2002}, we set up the system in a way similar to the expulsion study, and solve the passive $\psi$ scalar equation in the background of a stationary eddy. The $\psi$ field has a uniform gradient in the initial state: $\psi_0(x,y)=B_{\psi0}(x+L_0/2)$ where $B_{\psi0}$ is a coefficient analogous to the magnitude of the external magnetic field in MHD.

\begin{figure}[htbp] 
    \centering
    \includegraphics[width=\columnwidth]{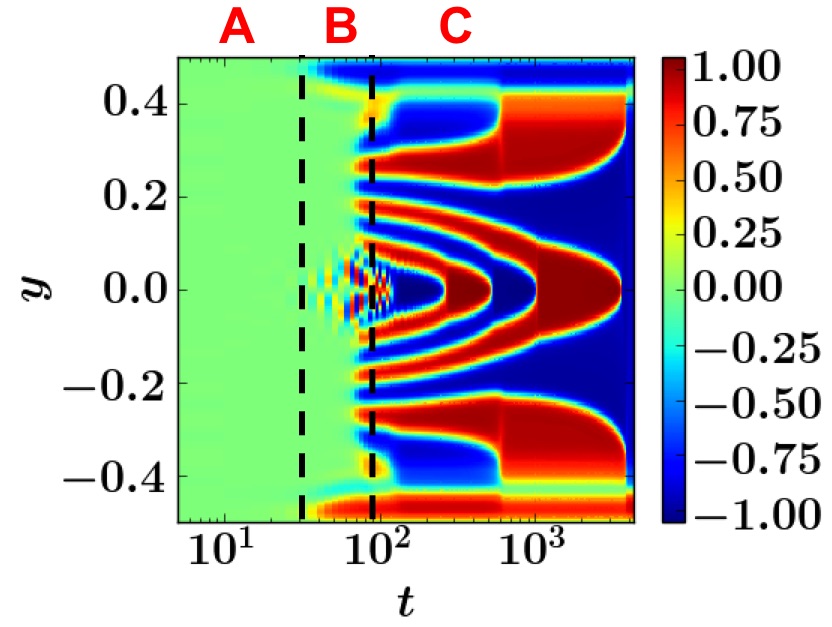}
    \caption{The evolution of $\psi$ at $x = 0$ with time. The three stages are distinguished by black dashed lines, and marked as A, B, and C, respectively. In the target pattern stage (C), the merger process is shown as the corner of the ``$>$'' shape. Reprinted with permission from Xiang Fan, P. H. Diamond, and L. Chac\'on, Phys. Rev. E \textbf{96}, 041101(R) (2017). Copyright 2017 American Physical Society.}
    \label{psi_evolution_target_shape}
\end{figure}

\begin{figure}[htbp] 
    \centering
    \includegraphics[width=\columnwidth]{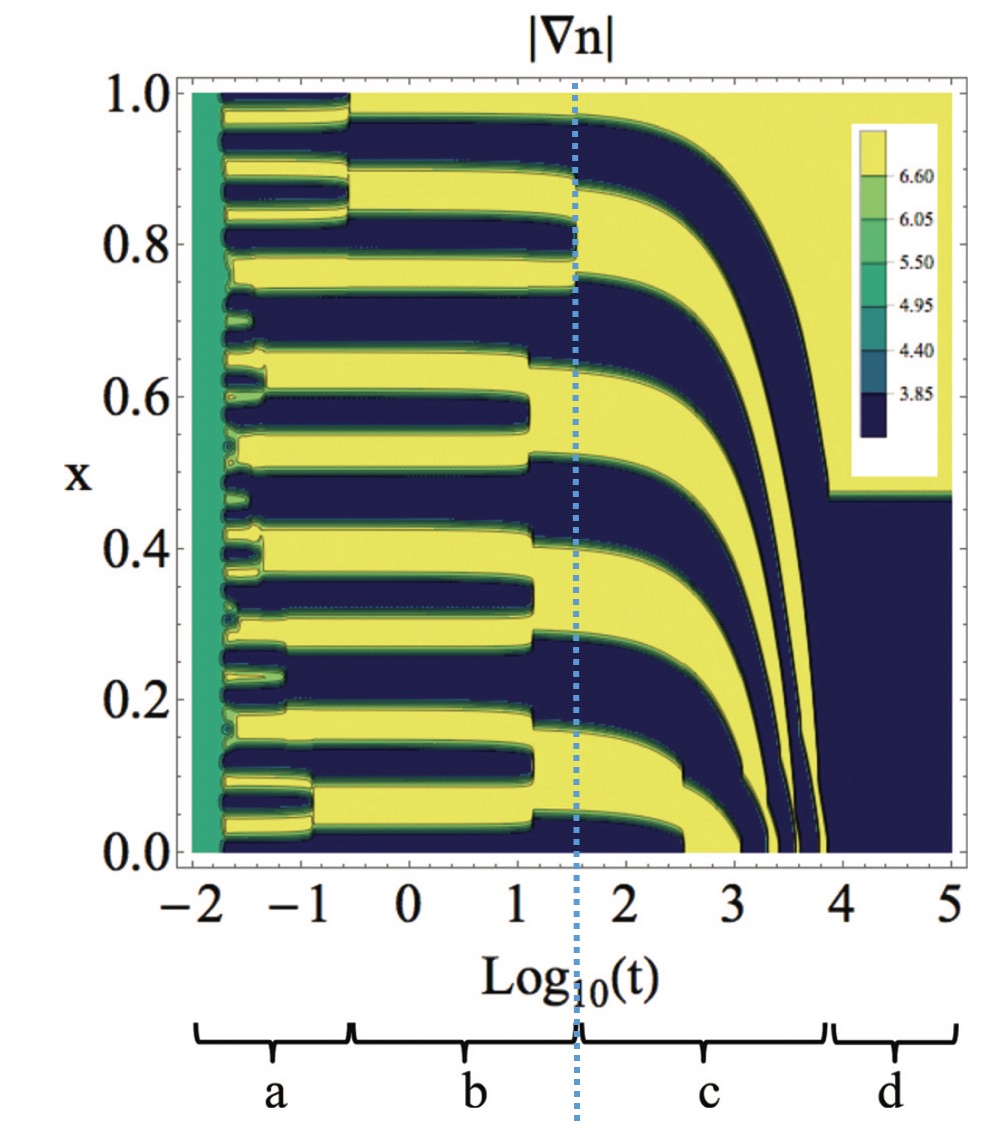}
    \caption{The staircase and step merger in confined plasma turbulence: contour plot of the time evolution of $|\nabla n|$ along the plasma radius. Different stages of evolution are: (a) Fast merger of micro-steps and formation of meso-steps. (b) Coalescence of meso-steps to barriers. (c) Barriers propagate along the gradient, condense at boundaries. (d) Stationary profile. Reprinted with permission from Ashourvan and Diamond, Phys. Rev. E \textbf{94}, 051202(R) (2016). Copyright 2016 American Physical Society.}
    \label{staircase}
\end{figure}

\begin{figure}[htbp] 
    \centering
    \includegraphics[width=\columnwidth]{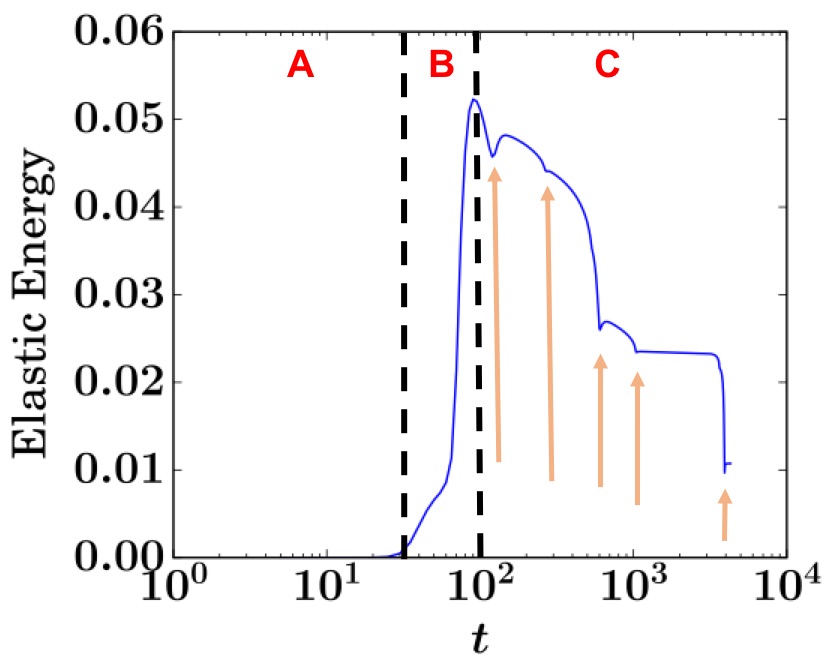}
    \caption{The time evolution of elastic energy. Note that logarithm scale is used for the t axis. A: the jelly roll stage; B: the topological evolution stage; C: the target pattern stage. The dips marked by orange arrows are due to band mergers. Reprinted with permission from Xiang Fan, P. H. Diamond, and L. Chac\'on, Phys. Rev. E \textbf{96}, 041101(R) (2017). Copyright 2017 American Physical Society.}
    \label{EB_t}
\end{figure}

Three stages are observed. There are: (A) an (initial) ``jelly roll'' stage, (B) the topological evolution stage, and (C) the target pattern stage. In the ``jelly roll'' stage, the stripes produced by spinodal decomposition (by negative diffusion) are wound up into spiral a shape. In the stage of topological evolution, the $\mathbf{B}_\psi$ lines ``reconnect'' and the spiral stripes evolve to concentric annuli, called a \textit{target pattern}. See Fig.~\ref{topology_change} for an illustration of the topological evolution. The target pattern is meta-stable, so this stage is long lived. The target bands merge on a time scale which is exponentially long, relative to an eddy turnover time. 

\begin{figure}[htbp] 
    \centering
    \includegraphics[width=\columnwidth]{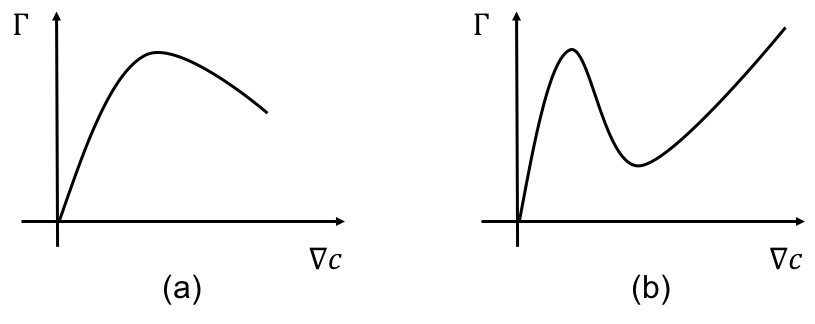}
    \caption{Flux-gradient relation.}
    \label{flux_gradient}
\end{figure}

The band merger process is shown in Fig.~\ref{psi_evolution_target_shape}. It is similar to the step merger in drift-ZF staircases as shown in Fig.~\ref{staircase} \cite{ashourvan_emergence_2017,ashourvan_how_2016}. Both merger phenomena progress by pairwise coalescence of stripes or steps, leading to progressive coarsening of the target or staircase patterns, as shown in Fig.~\ref{psi_fields} (e-h) and Fig.~\ref{psi_evolution_target_shape}. Note staircases occur in systems which exhibit a roll-over in the flux-gradient relation, i.e., Fig.~\ref{flux_gradient} (a)  \cite{balmforth_dynamics_1998,paparella_clustering_2012,ashourvan_emergence_2017,ashourvan_how_2016}; or bi-stability in that relation, i.e., Fig.~\ref{flux_gradient} (b). Flat regions (i.e., ``steps’’) form where flux $\Gamma$ increases rapidly with concentration gradient $-\nabla c$, and steep gradient regions (i.e., ``jumps’’) occur where $\Gamma$ is low. For either case, the crucial element is the presence of an interval of $\nabla c$ where the effective diffusivity $D_{eff}=-\delta\Gamma/\delta\nabla c$ goes negative, suggesting negative diffusion. Indeed, such a domain of negative diffusion seems unavoidable in a bi-stable system where $\delta\Gamma/\delta\nabla c$ is continuous. The dissipative operator on the RHS of the Cahn-Hilliard equation indeed exhibits a range of scales for which diffusion is negative. And, the range of the negative diffusion is ultimately limited by the stabilizing dominance of hyper-diffusion at smaller scales. Thus, there is a non-trivial similarity between the flux vs. gradient curve of systems in which staircase form, and the CH system. Indeed, reduced models of staircases and layering bear a resemblance to the CH equation.

The three stages are also reflected in the plot of energy v.s. time, shown in Fig.~\ref{EB_t}. In the ``jelly roll'' stage, the elastic energy remains small as compared to the later stages. Then in the topological evolution stage, the elastic energy rises. In the target pattern stage, the elastic energy decreases slowly and episodically. The band mergers are reflected in the plot as dips in the energy.

\begin{figure}[htbp] 
    \centering
    \includegraphics[width=\columnwidth]{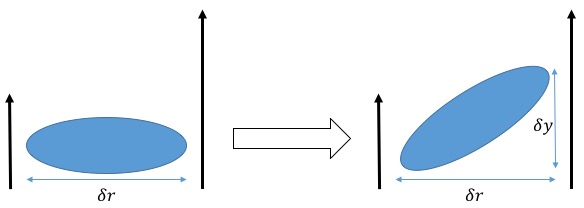}
    \caption{An eddy in a shear flow.}
    \label{shear}
\end{figure}

Analogous to the $\mathrm{Rm}^{1/3}$ time scale in MHD discussed in Ref.~\cite{rhines_how_1983}, the mixing time scale of the shear + dissipation hybrid case is $\tau_{mix}\sim\mathrm{Pe}^{1/5}\mathrm{Ch}^{-2/5}\tau_0$, where $\mathrm{Ch}\equiv\xi/L_0$ is the Cahn number, and $\mathrm{Pe}\equiv L_0 v_0/D$ is the Peclet number which is analogous to $\mathrm{Rm}$ in MHD. We speculate that this time scale represents the time for the topological change to occur. This expression is obtained analytically by considering the synergy of shearing and Cahn-Hilliard hyper-diffusion. Let $\delta r$ be the displacement in the radius direction, and $\delta y$ be the displacement along the flow, as illustrated in Fig.~\ref{shear}. We have $\frac{\intd}{\intd t}\delta y=s\delta r$, where $s$ is shear, thus $\langle\delta y^4\rangle\sim s^4\langle\delta r\rangle^4 t^4$. According to the Cahn-Hilliard Equation \ref{CHNS1}, and assuming the major process here is the hyper-diffusion, we have $\langle\delta r\rangle^4\sim D\xi^2t$, therefore $\langle\delta y^4\rangle\sim s^4 D\xi^2 t^5$. Let $\langle\delta y^4\rangle\sim L_y^4$ where $L_y$ is the scale of comparison, then the mixing time scale satisfies $1/\tau_{mix}\sim (s^4\frac{D\xi^2}{L_y^4})^{1/5}\sim\mathrm{Pe}^{-1/5}\mathrm{Ch}^{-2/5}/\tau_0$.

We also observe from simulations that the time to reach the maximum elastic energy is $\tau_m\sim\mathrm{Pe}\mathrm{Ch}^2\tau_0$, as shown in Fig.~7 in \cite{fan_formation_2017}. Clearly, single eddy mixing in CHNS is a multiple stage process, occurring on several time scales. Also, the formation of a meta-stable target pattern suggests a significant memory and resilience to mixing. These are due to the negative diffusion in the $\psi$ equation.

\section{Conclusion and Discussion}\label{conclusion_section}
In this paper, we have discussed the physics of the 2D CHNS system as a case study in elastic, active scalar dynamics. The comparison and contrast with 2D MHD is emphasized, as are the general lessons learned. For CHNS turbulence, the principle results of this paper are:
\begin{enumerate}
\item The CHNS system supports elastic waves (at $|\nabla\psi\neq0|$ interfaces), as well as eddies.
\item Blobs emerge at small scales, and merge to form fewer progressively larger structures. The blob scale evolves as $l\sim t^{2/3}$.
\item CHNS turbulence, with vorticity forced at large scale, manifests a dual cascade, with an inverse cascade of $\langle\psi^2\rangle$ and a forward cascade of enstrophy.
\item An elastic range is observed for $L_d<L<L_H$. Here $L_H\sim(\frac{\rho}{\xi})^{-1/3}\epsilon_\Omega^{-2/9}$ is the Hinze scale, at which the blob surface tension and fluid Reynolds stress are equal. $L_d$ is the dissipation scale. In the elastic range, the turbulence dynamics resemble those of MHD.
\item The $\langle\psi^2\rangle$ inverse cascade spectrum follows $\langle\psi^2\rangle_k\sim k^{-\alpha}$, with $\alpha=-7/3$, consistent with expectations based on scaling as in 2D MHD. The energy spectrum does not, but rather scales as $E_k\sim k^{-3}$, which resembles the spectrum in the enstrophy cascade range for a 2D fluid.
\item The resolution of the apparent paradox above is that, in the CHNS, elastic back reaction on the flow is restricted to the interfacial layers between blobs. As blobs coalesce, the number of blobs and the effective length of the interface decreases, thus weakening elastic back-reaction. In this limit, the dynamics approach that of a simple 2D fluid.
\end{enumerate}

For the study of single eddy mixing, the principal results are: 
\begin{enumerate}
\item The close analogy between the dynamics of mixing by a single eddy in kinematic 2D CHNS and flux expulsion in kinematic 2D MHD was noted and elucidated.
\item Episodic evolution of the elastic energy was observed, over multiple time scales. Elastic energy evolves through an initial ``jelly roll'' stage of wind-up, a fast topological stage of $\mathbf{B}_\psi$ reconnection, and a long time target pattern stage.
\item Target pattern formation was observed on long time scales. The bands of the target undergo pairwise mergers on time scales which are exponentially long in $\mathrm{Pe}$. This sequence of mergers resembles that between stages in a staircase.
\end{enumerate}

The most compelling topic for future work is the study of \textit{turbulent transport} mixing of mean concentration contrast ($\nabla\langle\psi\rangle$) in CHNS. The interesting question here is whether transport will be suppressed for large $\mathrm{Pe}(\nabla\langle\psi\rangle)^2$, much like it is for large $\mathrm{Re}\langle B\rangle^2$ in 2D MHD \cite{tobias_-plane_2007,cattaneo_effects_1994,cattaneo_suppression_1991,vainshtein_nonlinear_1992,guo_magnetic_2012}. This would suggest the existence of a self-stabilizing regime, where even strong stirring would not effectively mix a mean gradient. In MHD at large $\mathrm{Re}$, only a moderate $|\nabla\langle A\rangle|=|B_0|$ is required, as it is the \textit{small scale} magnetic fields which hold memory and inhibit mixing. Whether the analogy holds for the CHNS remains to be seek, as the ``Zeldovich relations'', connecting flux to intensity, are different for the two systems. However, note that the CHNS system manifests on intrinsic tendency to undo mixing by phase separation.

As a response to the broader questions question of what general lessons we learned in the course of this research, we return to the physics issues and challenges discussed in the introduction. This study has illuminated several aspects of active scalar turbulence beyond the confines of CHNS system. On the subject of dual cascades, this study illustrates that while multiple cascades can co-exist, some are more important than others. In CHNS, the evidence suggests that the inverse cascade of $\langle\psi^2\rangle$ (i.e. due blob coalescence) is the dominant process. Indeed, a major result of this work is the discovery that blob coalescence can modify the forward cascade by restricting elastic back reaction. Regarding ``blobby turbulence'', this study clearly demonstrates the utility of a real space approach. Here, real space is where we learn how blob structures modify the cascades. It also shows that the natural competition between eddy straining and droplet coalescence defines an important emergent scale, the Hinze scale. Similar scales are likely to emerge in other realizations of blobby turbulence. Finally, we learn that the negative diffusion leads to the formation of novel patterns in simple systems. Here, a good example is the target pattern formed in single eddy mixing. 

As to the question of ``what do we lean from all this?'', we offer the answers:
\begin{enumerate}
\item do not focus myopically on power laws! Real space quantities like packing fraction $P$ and interface structure are crucial to understanding the key dynamics.
\item one player in a dual cascade can modify or constrain the dynamics of the other.
\item somewhat contrary to conventional wisdom, the $\langle\psi^2\rangle$ inverse cascade is the robust nonlinear transfer process in CHNS. This raises the interesting question of what, really, is the essential process in MHD? It also suggests the question of whether 2D MHD turbulence can be approached as a competition between flux aggregation and fragmentation.
\end{enumerate}

More generally, the study supports the idea that exploring differences \textit{and} similarities between related, but distinct, systems is a useful approach to understanding turbulence in complex media.

\section{Acknowledgements}
We thank David Hughes and Annick Pouquet for interesting and useful conversations, Xiang Fan thanks Hui Li and Los Alamos National Laboratory for its hospitality and help with computing resources. We also thank participants in the 2015 and 2017 Festival de Theorie (Aix-en-Provence, France), for numerous stimulating interactions. This research was supported by the US Department of Energy, Office of Science, Office of Fusion Energy Sciences, under Award No. DE-FG02-04ER54738 and CMTFO Award No. DE-SC0008378.

\bibliographystyle{unsrt}
\bibliography{pop_invited} 

\end{document}